\documentclass[12pt,twocolumn,a4paper]{article}

\usepackage{times}
 \usepackage[square,sort,comma,numbers]{natbib} 
\usepackage{graphicx}

\usepackage{multirow}
\usepackage{tabularx} 
\title{Preliminary study for the measurement of the Lense-Thirring effect with the Galileo satellites}
\author{
Beatriz Moreno Monge, Rolf Koenig\thanks{Corresponding author. E-mail: rolf.koenig@gfz-potsdam.de}, Grzegorz Michalak
\\ \small{GFZ German Research Center for Geosciences (Germany)}
\\
Ignazio Ciufolini
\\ \small{Dipartimento di Ingegneria dell'Innovazione, Universit\'a  del Salento (Italy)}
\\
 Antonio Paolozzi, Giampiero Sindoni \\ \small{Scuola di Ingegneria Aerospaziale and DIAEE, Sapienza Universit\'a di Roma (Italy)} }

\begin{document}
\label{paper:myname}

\maketitle

\begin{abstract}
The precession of the orbital node of a particle orbiting a rotating mass is known as Lense-Thirring effect (LTE) and is a manifestation of the
general relativistic phenomenon of dragging of inertial frames or frame-dragging.~The LTE has already been measured by using the
node drifts of the LAGEOS satellites and GRACE-based Earth gravity field models with an accuracy of
about 10\% and will be improved down to a few percent
with the recent LARES experiment.~The Galileo system will provide 27 new node observables for the LTE estimation and their combination
with the LAGEOS and LARES satellites can potentially reduce even more the error
due to the mismodeling in Earth's gravity field.~However, the accurate determination of the Galileo orbits requires the
estimation of many different parameters, which can absorb the LTE on the orbital nodes.~Moreover, the accuracy of the Galileo orbits and hence,
 of their node drifts, is mainly limited by the mismodeling in the Solar Radiation Pressure (SRP).~Using
simulated data we analyze the effects of the mismodeling in the SRP on the Galileo nodes and
propose optimal orbit parameterizations for the measurement of the LTE from the future Galileo observations.
\end{abstract}

\section{Introduction}
In 1918, Lense and Thirring~\cite{LenseThirring1918} proved that a particle orbiting around a central body endowed with an angular momentum $J$ experiences
a nodal precession $\dot{\Omega}$ according to the expression

\begin{eqnarray}\label{lt}
\dot{\Omega} = \frac{2GJ}{c^2a^3 (1-e^2)^{3/2}}
\end{eqnarray}

where $a$ and $e$ are the orbital semimajor axis and eccentricity, $G$ is the gravitational constant and $c$ is the speed of light.

\indent One of the main error sources in the estimation of the Lense-Thirring effect (LTE) by using the nodes of an Earth satellite arises from the
uncertainties in the Earth's gravity field model.~In particular, the largest errors  proceed from uncertainties in the
first even zonal harmonics of a spherical expansion of the Earth potential, i.~e., $J2$, $J4$, $J6$, ..., and their variations in time.

\indent The combination of two node observables  in order to remove the influence of the first even
zonal harmonic, $J2$, on the node drift was first proposed in~\cite{Ciufolini1986}.~Then, the two laser-ranged LAGEOS satellites and the high-accurate
Earth gravity field models based on GRACE observations (f.~i., EIGEN-GRACE02S, EIGEN-GRACE03S, JEM03G), have provided for the measurement of the LTE
with an accuracy
of about 10\%~\cite{Ciufolini2011}.~The combination with the new LARES satellite~\cite{website:LARES}  will allow to eliminate the
uncertainties due to both, $J2$ and $J4$, thus being possible to obtain a measurement of the LTE with an accuracy of the order of 1\%~\cite{Ciufolini2011}.

\indent The Galileo system will provide a new node observable from a total of 27 satellites, whose combination with the LAGEOS and LARES satellites
will potentially reduce the uncertainty due to the mismodeling in the Earth gravity field.~However, there are two issues that greatly impact the
LTE measurement with the Galileo satellites.~The first of those is that
the accurate determination of the Galileo orbits and their node drift requires the estimation of many different parameters, such as initial state vectors,
empirical accelerations, clock offsets, station coordinates, etc, which can absorb partially or completely the LTE.~The second issue is that
the final accuracy of the estimated Galileo orbits is mainly limited by the mismodeling in the Solar Radiation Pressure (SRP), which constitutes the primary error
source in the determination of the LTE with the Galileo nodes.

\indent In the present work, the effects of both, the mismodeling in the SRP and the orbit parameterization, on the Galileo node drift determination are analyzed
by means of simulated Galileo orbits and observations.

\section{Simulation of Galileo orbits and observations}\label{simulation}
A set of Galileo orbits and observations corresponding to days 1-4 of January, 2008 are simulated and used in a series of different numerical tests.~In
the simulation, the EPOS-OC software (Earth Parameter and Orbit System– Orbit Computation~\cite{Zhu2004}) is used.

\indent The Galileo orbits are simulated according to the specifications given by ESA~\cite{website:ESA}, a Walker
27/3/1 configuration with a semi-major axis of 29600 km, an inclination of 56 deg, zero eccentricity, a separation between
planes equal to 120 deg and a change of mean anomaly for equivalent satellites in neighbouring planes of 13.3
deg.~The main physical background models used in the simulations are presented in Table~\ref{tab:physicalmodels}.~The corresponding datasets of
Galileo code and phase observations for a global network of 80 stations are obtained.~In the data simulation and in the subsequent orbit recovery,
identical background models are used  and hence, the errors due to uncertainties in these models are not addressed in this work.

\indent In EPOS-OC the SRP acceleration $\ddot{\mathbf{r}}$ is computed by means of the expression

\begin{equation}\label{eq:acc}
 	\ddot{\mathbf{r}} = \left( \frac{A}{R}  \right)^2 \frac{1}{m} F_{rad} \frac{\mathbf{R}}{R}
\end{equation}

where $A$ represents the Astronomical Unit (AU) in meters, $m$ is the satellite mass, $\mathbf{R}$  is the heliocentric vector pointing out to the satellite with
 module $R$ and $F_{rad}$ is the direct pressure force computed by means of a model.~For the time being, the SRP models implemented in EPOS-OC
are the ROCK4 model~\cite{Fliegel1992} for GPS-type satellites and the cannon ball or macro models dedicated to specific non-GPS satellites
 (f.~i.,~\cite{Marshall1992}).~In our simulations the Galileo satellites were considered to be GPS-Block-II-like satellites and thus, the ROCK4 model
 was applied.~The ROCK4 model is recommended by the IERS conventions~\cite{McCarthy1992}  for the modeling of the SRP effect.~It includes
the dimensions and optical
properties of the GPS spacecraft surfaces.

\begin{table}[h]
\centering \small
\begin{tabular}{|l|l|} \hline
Gravity field & EIGEN-6C 12 x 12 \\
 & \cite{Forste2011} \\ \hline
Earth tide & IERS Conv.~\cite{PetitLuzum2010} \\  \hline
Ocean tide & EOT11a~\cite{Savcenko2011} \\  \hline
Atmospheric tide &  Biancale-Bode~\cite{Biancale2006} \\  \hline
Lunisolar and planetary & JPL DE421 ~\cite{Folkner2009} \\
 perturbations   & \\ \hline
Ocean pole tide & Desai~\cite{Desai2002} \\  \hline
Earth Orientation & EOP08C04 \\
  Parameters & \\ \hline
Nutation and precession & IERS Conv.~\cite{PetitLuzum2010} \\ \hline
Earth albedo & Analytic model  \\
&  by Heurtel \\\hline
\end{tabular}
\caption{Physical background models used in the simulation of the Galileo orbits and observations}
\label{tab:physicalmodels}
\end{table}

\indent In order to account for  deficiencies in the SRP model, the Eq.~(\ref{eq:acc}) is multiplied
by a global scaling factor (F0 hereafter) and different parameters can also be added, like global biases in the X, Y
and Z directions in a satellite body-fixed reference system.~The influence of the Earth's and Moon's
shadow is also considered.

\indent The creation of a more accurate SRP model for Galileo will require the knowledge of the vehicles characteristics in
terms of shape, size, weight and surface optical properties.~Then, a macro model
could be adopted for the Galileo SRP computation, along with an appropriate attitude model which also accounts for the  yaw turns
during satellite midnights (shadow turns) and noons. As of today, little information has been published about
the characteristics of the Galileo satellites. Some general features taken from~\cite{website:ESA} are compiled in Table~\ref{tab:galchar}.

\begin{table}[h]
\centering
\begin{tabular}{|l|l|} \hline
Bus dimensions & 2.7 x 1.1 x 1.2 m$^3$ \\  \hline
Solar array span & 13 m \\  \hline
Mass & 700 Kg \\  \hline
\end{tabular}
\caption{Galileo satellite features}
\label{tab:galchar}
\end{table}

\section{Optimal parameterizations for the LTE estimation}\label{param}
\begin{figure}[ht]
\centering
\includegraphics[width=1\linewidth]{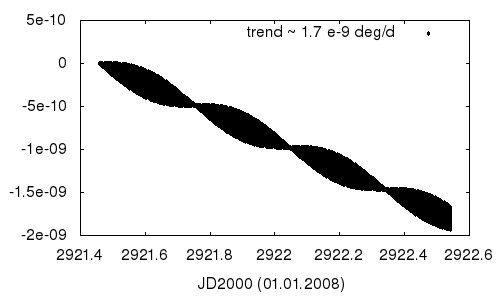}
\caption{Orbital node differences (deg) between two sets of simulated Galileo orbits with and without LTE.}\label{fig:fig1}
\end{figure}

 According to Eq.~(\ref{lt}), the nodal precession of the Galileo satellites due to the LTE is $1.7 \cdot 10^{-9}$ deg/d.~This
node drift holds for all 27 Galileo satellites of the full constellation. Figure~\ref{fig:fig1} shows the comparison of two sets of
\textit{simulated} orbits over a period of 26 h generated with and without modeling of the LTE.

\indent To measure the LTE from Galileo observations the nodal drift due to the LTE must be present in the node positions \textit{estimated}
in a precise orbit determination process where the LTE is not modeled.~This means that the estimated node positions without LTE modeling must differ from the true node positions and
this difference must not be absorbed by  the estimated parameters.~In
order to find out under which circumstances this holds, a set of 26 h of Galileo orbits and noise-free code and phase observations are
simulated including the modeling of the LTE.~Then, the simulated observations are used to recover the Galileo orbits without modeling the LTE, thereby estimating
the following parameters:

\begin{itemize}
\item[-] Initial orbital elements for each satellite (position and velocity)
\item[-] Empirical accelerations in the along-track and normal directions (4 coefficients per arc)
\item[-] SRP F0, Y- and Z-bias
\item[-] Earth's albedo scaling factor
\item[-] Station coordinates
\item[-] Tropospheric delays (10 per satellite-station pair)
\item[-] Phase ambiguities
\item[-] Clock offsets
\end{itemize}

\indent The \textit{estimated} (without LTE) and \textit{simulated} (with LTE) node positions are compared and the differences
analyzed in order to identify the parameters absorbing the LTE.~An optimal parameterization for the LTE measurement must allow
to observe the node drift when comparing the estimated and simulated orbits.

\begin{figure*}[t]
\centering
\includegraphics[width=1\linewidth]{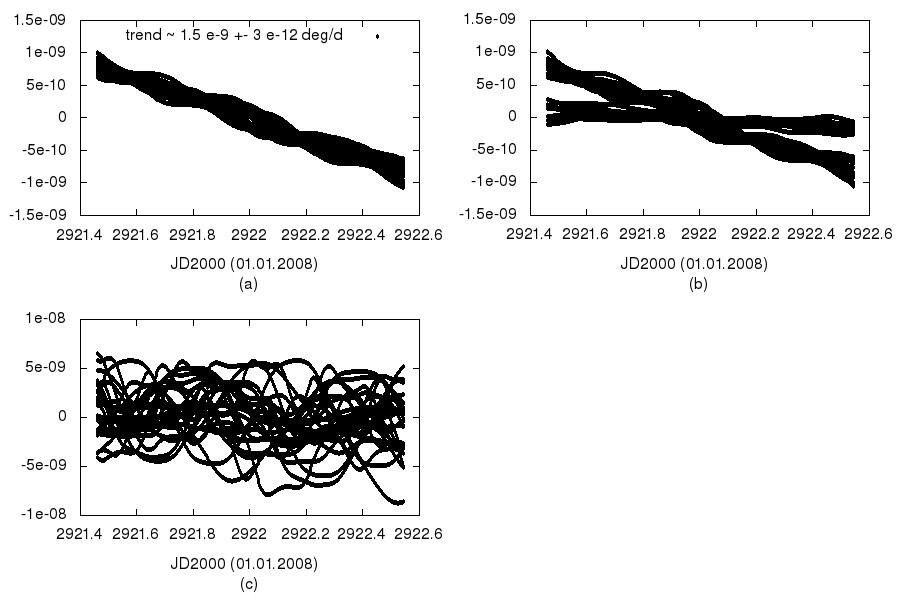}
\caption{Orbital node  differences (deg) between estimated (without LTE) and simulated (with LTE) Galileo orbits when (a) station coordinates are constrained with NNTRS conditions and the empirical
accelerations in the orbit normal direction are fixed (optimal parameterization), (b) the X-bias is additionally estimated and (c) the optimal parameterization is applied to
 noisy observations.}\label{fig:fig234}
\end{figure*}

\indent It was found that the free estimation of the station coordinates introduces large errors in the node positions of the order of $10^{-8}$ deg
and, as a consequence, the station coordinates must be either fixed or highly constrained, what can be done by imposing a set of
No-Net-Translation-Rotation-Scale (NNTRS) conditions on the whole ground network.~In addition, the estimation of empirical
accelerations in the normal direction to the orbital plane absorbs part of the LTE and  consequently, their estimation must be avoided,
 with a concurrent loss in orbit recovery accuracy.~Provided that these two conditions are fulfilled the LT signal is
observed in the differences between the estimated and simulated
nodes positions, like in Fig.~\ref{fig:fig234}~(a).~A linear regression of the node position differences provides an estimate of the LTE drift of approximately
 $1.5 \cdot 10^{-9}$ deg/d, close to the expected value, $1.7 \cdot 10^{-9}$ deg/d, with a standard deviation of $3 \cdot 10^{-12}$ deg/d and a small post-regression RMS of $9 \cdot 10^{-11}$ deg.
In conclusion, this parameterization seems to be optimal for the LTE estimation from 1-day Galileo arcs in the sense
that the LTE is not absorbed by the estimated parameters. 

\indent An additional test has been performed by estimating another parameter of the SRP model, the X-bias, in addition to F0, Y- and Z-bias.
The differences in the node positions are presented in Fig.~\ref{fig:fig234}~(b).~In this case the LT signal has been
absorbed for some satellites and hence, the estimation of the SRP X-bias must be avoided.~This can be safely done, since the estimation of the X-bias does not
introduce a significant improvement in the accuracy of the recovered orbit and therefore is usually dismissed.


\indent The results obtained when applying the
 optimal parameterization to noisy observations are also analyzed.~For this purpose, a Gaussian noise has been
introduced to the simulated observations with standard deviations of 50 cm for code and 3 mm for phase ranges. The differences between
estimated and simulated node positions are shown in Fig.~\ref{fig:fig234}~(c).~In this case, the node drifts due to the LTE is
hidden behind the large noise.~In fact, the RMS of the node differences is about $2.7 \cdot 10^{-9}$ deg,
quite larger than the expected node displacement from the LTE for 1 day, i.~e., $1.7 \cdot 10^{-9}$ deg/d.~Therefore, in order to obtain a more precise
estimation of the LTE, longer arcs shall be used, f.~i.~with 3-day arcs the LTE amounts to $5.1 \cdot 10^{-9}$ deg/d, which is significantly larger than
the noise mentioned before.

\begin{figure*}[t]
\centering
\includegraphics[width=1\linewidth]{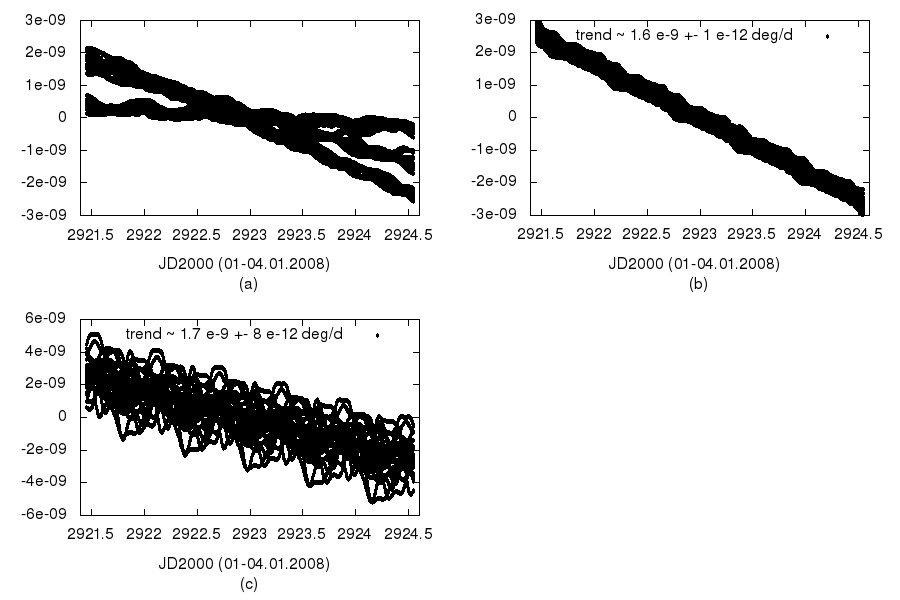}
\caption{Orbital node differences (deg) between estimated (without LTE) and simulated (with LTE) 3-day Galileo orbits when (a) the optimal parameterization
for 1-day arcs is applied, (b) the Earth albedo is additionally fixed (optimal parameterization for 3-day arcs) and
(c) the optimal parameterization for 3-day arcs is applied to noisy observations.}\label{fig:fig567}
\end{figure*}

\indent Thus, for the next tests, 3-day Galileo orbits and noise-free observations are simulated with modeling of the LTE. Then, the Galileo orbits
are estimated without modeling the LTE
from the observations by applying the optimal parameterization
 for 1-day arcs.~The same number of empirical coefficients are estimated for the 3-day arcs as for the 1-day arcs.~The differences
between the estimated and simulated node positions are presented
in Fig.~\ref{fig:fig567}~(a), where it can be observed that the LT signal is absorbed for some satellites and hence, the optimal parameterization for 1-day
arcs is not suitable for 3-day arcs.

\indent The parameters absorbing the LTE are identified as the SRP Y-bias and the Earth's albedo scaling factor.~Fixing either of
them results in a perfect recovery of the LTE,
as shown in  Fig.~\ref{fig:fig567}~(b).~Nevertheless, fixing the Y-bias to an incorrect value would introduce large errors
in the node positions and has to be avoided.~On the contrary, the influence of Earth's albedo on the Galileo orbits is very small
and therefore can safely be neglected.~Consequently, the optimal parameterization
for 3-day arcs coincides with that for 1-day arcs with the exception of the Earth's albedo scaling factor, which must be fixed for the 3-day arcs.



\indent Finally, the optimal parameterization for 3-day arcs is applied to noisy observations, the differences in the node positions are
in Fig.~\ref{fig:fig567}~(c).~The linear regression provides a trend of $1.7 \cdot 10^{-9}$ deg/d, with 
an RMS of $1.0 \cdot 10^{-9}$ deg, 5 times smaller than the total displacement of the Galileo nodes due to the LTE
after 3 days.~In conclusion, the 3-day arcs in combination with the optimal parameterization proposed here seem suitable for the precise measurement of
the LTE from
noisy Galileo observations in absence of other modeling errors.~Then, using real Galileo observations and estimated orbits, the LTE will
 be measured as the difference between the estimated node positions in the overlap of two consecutive Galileo orbital arcs.~Assuming a precision
of the estimated node positions
at the level of the RMS, this is $1.0 \cdot 10^{-9}$ deg for 3-day arcs, the combination of 27 node observables to compute the difference in the
overlap yields a precision
of $1.0 \cdot 10^{-9}  \cdot \sqrt{2}/\sqrt{27} = 0.27 \cdot 10^{-9}$ deg for the LTE estimation.~Thus, a precision of about 5\% of the LTE would be
reached with two 3-day Galileo
 arcs.~Similarly, the average of LTE estimations obtained in a series
of $n$  overlaps increases the precision by a factor
$1/\sqrt{n}$.~Thus, in order to achieve a precision in the LTE estimation of 1\%, a minimum of 30 Galileo 3-day arcs need to be processed.


\begin{figure}[t]
\centering
\includegraphics[width=1\linewidth]{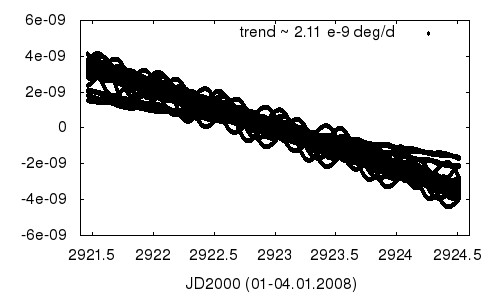}
\caption{Orbital node  differences (deg) between estimated 3-day GPS orbits with and without modeling the LTE when the optimal parameterization is applied.}\label{fig:fig8}
\end{figure}

\indent The optimal parameterization has been verified with real GPS data through the estimation of a 3-day GPS
 orbit in two different cases: with and without modeling of the LTE.~The differences in the node positions between the two sets of estimated
orbits are in Fig.~\ref{fig:fig8} and show a drift of the GPS nodes
of about  $2.1 \cdot 10^{-9}$ deg/d, close to the theoretical value of $2.3 \cdot 10^{-9}$ deg/d.~This
means that this optimal parameterization allows to observe the LTE in the GPS nodes provided that there are no errors in the background models and
the nodes are accurately estimated to the few $10^{-9}$ deg level.~This however is difficult to achieve mainly due to SRP modeling errors.

\section{Effects of the SRP on the Galileo nodes}\label{SRP}
\begin{figure*}[t]
\centering
\includegraphics[width=1\linewidth]{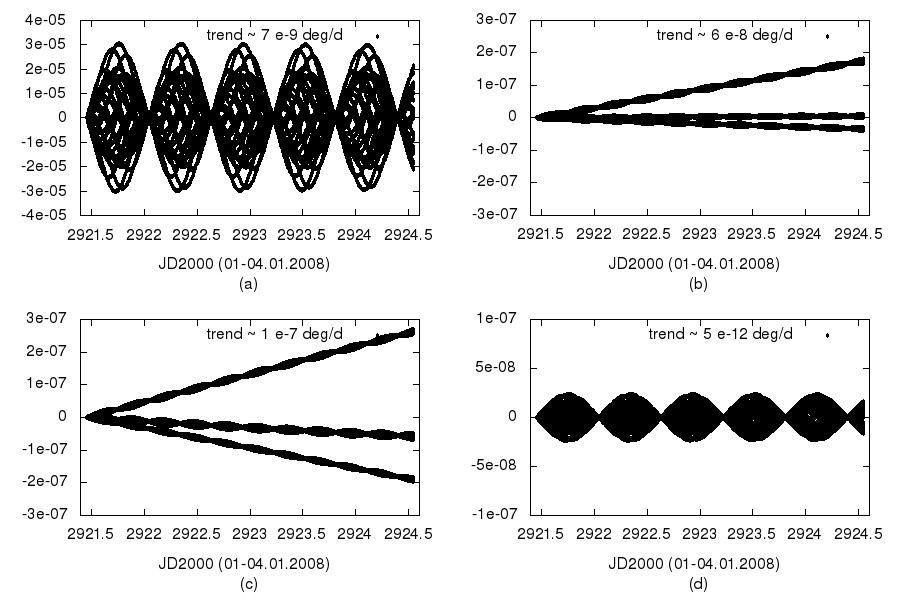}
\caption{Effect of the SRP on the Galileo nodes  of simulated orbits (deg) when the following parameters are considered: (a) F0 only, (b) X-bias only,
(c) Y-bias only and (d) Z-bias only.}\label{fig:fig910}
\end{figure*}

In this section we analyze the effects of the SRP on the Galileo nodes and the errors in the \textit{estimated} nodes  due to
 mismodeling of SRP.~For that purpose, a set of Galileo orbits is simulated
by including various parameters of the SRP model, then they are compared to a set simulated without
SRP model.~In first place a set of Galileo orbits with F0 equal to 1 and X-, Y- and Z-biases equal to 0 is tested.~The
effect is shown in Fig.~\ref{fig:fig910}~(a).~A periodical displacement of the node
 is observed with an amplitude of up to $3 \cdot 10^{-5}$ deg and a period equal to a complete revolution of the Galileo satellites.~Moreover
the SRP introduces a drift of $7 \cdot 10^{-9}$ deg/d, about twice the level of the LTE.

\indent The effects due to the presence of the X-, Y- and Z-bias (all simulated
with magnitude $10^{-10}$ m/s$^2$)
are shown in Figs.~\ref{fig:fig910}~(b), (c) and (d) respectively.~It can be observed that the X- and Y-bias yield a significant trend  of
up to $6\cdot 10^{-8}$ deg/d  and up to $1\cdot 10^{-7}$ deg/d respectively, in both cases depending on the orbital planes.~Conversely, the Z-bias produces a
periodical displacement of the node
with an amplitude of $2.5 \cdot 10^{-8}$ deg and a period of one revolution.~The resulting node drift due to the Z-bias is $5\cdot 10^{-12}$ deg/d only.~In
summary, the SRP parameters must be handled carefully,
since they can introduce large errors in the node thus easily masking the sought for LTE.

\begin{figure*}[t]
\centering
\includegraphics[width=1\linewidth]{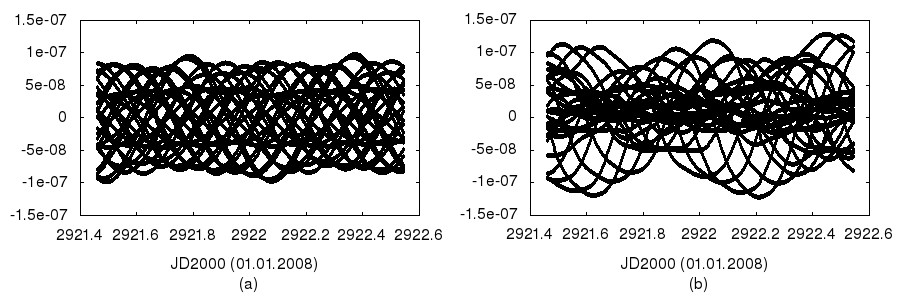}
\caption{Orbital node differences (deg) between estimated (without LTE) and simulated (with LTE) Galileo orbits (a) with an error in F0 of 20\% and
(b) with an error in F0, Y- and Z-bias.}\label{fig:fig1112}
\end{figure*}

\indent As a next step, the error in the \textit{estimated} nodes due to the mismodeling in the SRP is analyzed in two different cases.~In the
first case 1 day of Galileo orbits and observations are simulated with F0 = 1 and the orbits are recovered from the observations by fixing F0 = 1.2.~This
represents a deliberate error of 20\% of the full SRP model coming from a mismodeling in the satellite area,
mass, surface properties, etc.~In GPS orbit determination F0 varies periodically ranging from 0.5 to 1.5 or even more and hence, an error of 20\%
seems realistic.~No X-, Y- and Z-biases are considered in this first case and the optimal parameterization for 1-day arcs is used in the recovery.~Like
in the previous tests
the LTE is introduced in the data simulation but it is not modelled in the recovery of the orbits.~The
differences in the node between estimated and simulated orbits are
 presented in Fig.~\ref{fig:fig1112}~(a).~A
periodical error with a magnitude of up to  $1\cdot 10^{-7}$ deg is clearly observed,  the node drift due to the LTE is not visible.~As
a consequence, the global scaling factor of the SRP model shall not be fixed or highly
constrained, but rather estimated.~The estimation of F0 absorbs the deficiencies in
the SRP model and does not affect the measurement of the LTE, as shown in Section~\ref{param}.

\indent In a second case, the Galileo orbits and observations are simulated by including Y- and Z-biases in the SRP model and an increasing F0 from the beginning to
the end of the arc, according to the values
 given in Table~\ref{tab:srppar}.~These values are realistic since they are obtained from the determination of GPS orbits.~In the subsequent estimation of the
Galileo orbits, F0 is fixed to 1 and the Y- and Z-biases are fixed to 0.~The differences between estimated and simulated nodes are
 shown in Fig.~\ref{fig:fig1112}~(b).~This time, the errors observed in the estimated nodes reach more than $1\cdot 10^{-7}$ deg mainly due to the mismodeling
of the Y- and Z-bias,
 the LT node drift is not visible.~Thus, the Y- and Z-bias of the SRP shall not be fixed but rather estimated since fixing them to incorrect values introduces errors in the node position
two orders of magnitude larger than the LTE.

\begin{table}[h]
\centering
\begin{tabular}{|l|l|} \hline
parameter &
value \\ \hline
F0 &
0.999 - 1.001 \\
 & (total variation = 0.002) \\ \hline
Y-bias &
-0.59600E-10 \\ \hline
Z-bias &
-0.13400E-07 \\  \hline
\end{tabular}
\caption{SRP parameters as obtained from GPS orbit determination and used in the simulations}
\label{tab:srppar}
\end{table}


\indent Finally, to analyze the effects of the SRP depending on the satellite surface properties,
a test is performed by using a macro model for the Galileo satellites.~A macro model (sometimes also called box-wing model in the literature)
takes into account the size and reflection properties of each surface of the satellite.~The Galileo attitude model is neglected here.~It
must be stressed however, that the attitude model is critical for the computation
 of the SRP effects on the Galileo orbits, since their panels are continuously reoriented to face the Sun.~Thus, the results
obtained here are a simplification of the real behaviour of the Galileo nodes.

\begin{figure}[t]
\centering
\includegraphics[width=1\linewidth]{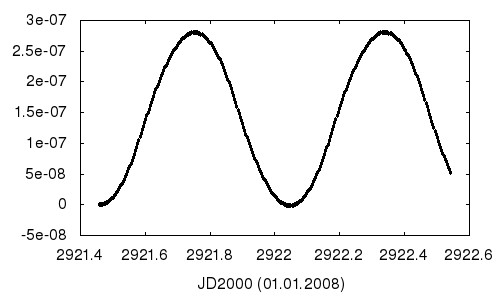}
\caption{Orbital node differences (deg) between simulated orbits considering gold or silver/aluminium coatings.}\label{fig:fig13}
\end{figure}

\indent  According to different
illustrations of the Galileo satellites (e.~g.~\cite{website:ESA} or~\cite{website:OHB}), two different sets of parameters are
set up, corresponding to two different satellite coatings, a gold coating and
a silver (similar to aluminium) coating.~The parameters of the macro models used are summarized in
 Table~\ref{tab:macro}.~It can be noted that the differences of the coefficients due to the choice of the coating amounts to 25\% already.~This
will directly transfer into the SRP modeling and corresponds roughly with the uncertainties of the SRP model used in the above.~The
effect on the nodes  due to the choice between the two different coatings can be observed
in Fig.~\ref{fig:fig13}.~The node differences are tremendous reaching $2.8 \cdot 10^{-7}$ deg peak to peak.~In the
end however, in Galileo orbits estimation the error due to a wrong choice of the surface coating will be reduced by the estimation of certain SRP parameters.

\begin{table}[h]
\centering \footnotesize
\begin{tabular}{|l|c|c|c|c|c|} \hline

\multirow{2}{*}{Surface} & Area    &  \multicolumn{4}{c|}{Refl. coeff. visible}  \\ \cline{3-6}
                         & (m$^2$) & \multicolumn{2}{c|}{gold} & \multicolumn{2}{c|}{silver}    \\ \hline
                         &         & geom & diff               & geom & diff \\ \hline

bus top     & 1.32 & 0.14 & 0.56 & 0.18 & 0.72   \\ \hline
bus bottom  & 1.32 & 0.14 & 0.56 & 0.18 & 0.72    \\ \hline
bus left    & 2.75 & 0.14 & 0.56 & 0.18 & 0.72    \\ \hline
bus right   & 2.75 & 0.14 & 0.56 & 0.18 & 0.72    \\ \hline
bus front   & 3.00 & 0.14 & 0.56 & 0.18 & 0.72   \\ \hline
bus back    & 3.00 & 0.14 & 0.56 & 0.18 & 0.72   \\ \hline
panel left  &11.70 & 0.04 & 0.16 & 0.04 & 0.16    \\ \hline
panel right &11.70 & 0.04 & 0.16 & 0.04 & 0.16      \\ \hline

\end{tabular}
\caption{Macro model parameters for the Galileo satellites based on gold and silver coatings.}
\label{tab:macro}
\end{table}

\indent In conclusion, the measurement of the LTE with the Galileo nodes requires an accurate SRP model for what it is essential to know the Galileo
vehicles shape and surface
properties.~Also some model parameters such as F0, the Y- and Z-biases need to be estimated.~The
uncertainty arising from the SRP model is unavoidable, however a significant reduction could be
achieved only if the Galileo satellites would be equipped with accelerometers, which
measure all non-conservative forces acting on the satellites.


\section{Discussion and Conclusions}\label{conclusion}
Using simulated Galileo orbits and observations, the effect of the orbit parameterization on the detection of the LTE and
 the effects of  SRP on the Galileo nodes are analyzed.

\indent The best
parameterizations for the LTE estimation with 1-day and 3-day Galileo orbits are proposed.~It is shown that the best parameterization depends
on the arc length.~In general, the station coordinates cannot be freely estimated but they must be fixed or highly constrained, f.~i.~by
imposing a set of NNTRS conditions.~In addition, the empirical accelerations in the normal direction to the orbital plane and the
X-bias of the SRP model must not be estimated, since they absorb the LT signal in the orbital nodes.~When
processing 1-day arcs, the SRP parameters (F0, Y- and Z-bias) and the Earth's albedo scaling factor can simultaneously be estimated, however with 3-day arcs
the Earth albedo must be fixed.~These optimal parameterizations allow to estimate the LTE from \textit{noise-free} observations
provided that there are no errors in the background models.~The possible errors due to the mismodeling of the background
models shall be analyzed in future tests.



\indent The 3-day arcs in combination with the optimal parameterization seem suitable for a precise estimation of the LTE from \textit{noisy} Galileo
observations,
 since the noise in the estimated nodes
 is 5 times smaller than the total displacement of the node due to the LTE after 3 days.~In the future this must be confirmed with real Galileo
observations.~Then, assuming no errors in the background and SRP models, the LTE seems to be estimable from real Galileo data with
a precision of 1\% by using the nodes of 27 satellites
 and a minimum of 30 3-day arcs, again assuming no errors in the background models.

\indent The SRP introduces large periodical displacements on the
Galileo nodes, four orders of magnitude larger than the LTE.~The deviations
mainly cancel out after each complete revolution.~The resulting node drift is twice the value of the LTE.~In addition,
 the presence of X- and Y-biases produces significant nodal drifts, two orders of magnitude larger than the LTE depending on the orbital plane.~The Z-bias
introduces small periodical effects and an insignificant node drift.~Fixing the SRP parameters (F0, Y- and Z-bias) to incorrect values yields large
errors in the estimated nodes. Hence, these parameters shall
not be fixed or highly constrained in the estimations, allowing them to absorb the deficiencies in the SRP model.~The
choice of wrong reflectivity coefficients of the surface of the
satellites can give place to deviations of the nodes of up to two orders of magnitude larger than the LTE, which in the end can be
reduced by the estimation of certain SRP parameters.


%

\indent As a consequence, the measurement of the LTE by using the Galileo nodes requires an  accurate SRP model, like f.~i.~a macro model fitted to
the shape, size, weight and surface properties of the Galileo satellites. The macro model needs to be accompanied by an appropriate attitude model.~The
SRP parameters shall be estimated
(F0, Y- and Z-bias), allowing them to absorb the deficiencies of the SRP model. 

\section*{Acknowledgements}
The authors wish to thank the European Space Agency for supporting this research under the ESTEC project
"Space Tests of General RElativity Using the GAlileo Constellation (REGAL)", contract No. 4000103504/2011/NL/WE.

\bibliographystyle{plain}
\bibliography{moreno_LT_galileo}

\end{document}